# Does society show differential attention to researchers based on gender and field?

Sara M. González-Betancor [1] and Pablo Dorta-González [2,*]

[1] Department of Quantitative Methods in Economics and Management, Campus de Tafira, University of Las Palmas de Gran Canaria, 35017 Las Palmas de Gran Canaria, Spain; sara.gonzalez@ulpgc.es

[2] Institute of Tourism and Sustainable Economic Development (TIDES), Campus de Tafira, University of Las Palmas de Gran Canaria, 35017 Las Palmas de Gran Canaria, Spain

[*] Correspondence: pablo.dorta@ulpgc.es

**Abstract**

While not all researchers prioritize social impact, it is undeniably a crucial aspect that adds significance to their work. The objective of this paper is to explore potential gender differences in the social attention paid to researchers and to examine their association with specific fields of study. To achieve this goal, the paper analyzes four dimensions of social influence and examines three measures of social attention to researchers. The dimensions are media influence (mentions in mainstream news), political influence (mentions in public policy reports), social media influence (mentions in Twitter), and educational influence (mentions in Wikipedia). The measures of social attention to researchers are: proportion of publications with social mentions (social attention orientation), mentions per publication (level of social attention), and mentions per mentioned publication (intensity of social attention). By analyzing the rankings of authors - for the four dimensions with the three measures in the 22 research fields of the Web of Science database- and by using Spearman correlation coefficients, we conclude that: 1) significant differences are observed between fields; 2) the dimensions capture different and independent aspects of the social impact. Finally, we use non-parametric means comparison tests to detect gender bias in social attention. We conclude that for most fields and dimensions with enough non-zero altmetrics data, gender differences in social attention are not predominant, but are still present and vary across fields.

## Introduction

Governments are increasingly pushing researchers towards activities with social impact, including economic, cultural, and health benefits (Thelwall, 2021). Therefore, since the introduction of the term 'altmetrics' in 2010 (Priem et al., 2010), theoretical and practical research has been conducted in this discipline (Sugimoto et al., 2017). Altmetrics provide insight into how the results of scientific research are perceived and commented on beyond academia (Dorta-González and Dorta-González, 2022).

The pioneering work of Eysenbach (2011) found that tweets about scientific papers reflect their social impact and citizen attention. Bormann (2014) and Piwowar (2013) share this view, highlighting the ability of digital platforms to engage different publics and audiences. However, recent discussions have questioned whether altmetric indicators are an accurate measure of social influence. While they reflect some forms of attention (Thelwall, 2021), their validity for measuring social impact is uncertain (Bornmann et al., 2019).

To improve the accuracy of social impact measurement, Díaz-Faes et al. (2019) proposed a novel approach that combines bibliometric and altmetric methods, thus introducing a second generation of social media metrics. Advances in the field, such as the emergence of Twitter-based topic-actor networks as explored by Hellsten and Leydesdorff (2020), indicate a noticeable shift towards interactive and network-oriented perspectives in the study of social media metrics. In this evolving landscape, it is crucial to distinguish between primary metrics, which are related to the research objects themselves, and secondary metrics, which are associated with social media entities or users. These innovative methodologies provide a robust basis for potentially classifying users based on their behavior on social media platforms and facilitate comparative analysis among them, thereby enabling effective benchmarking. However, more research on altmetrics is needed, especially on the different types of social attention and their meaning. In this sense, each type of digital platform where a scientific article is mentioned reflects a different type of influence than the academic one, making it possible to quantify the task of transferring knowledge to society in several of its dimensions (Arroyo-Machado et al., 2022).

Studies of gender and altmetrics can inform efforts to promote gender equity in academia and scholarly communication. A paper exploring the gender gap in altmetric mentions found that online science dissemination is male-dominated and that female scientists are less likely to be in the top 25% of the most successful scholars online in all research fields studied (Vásárhelyi et al., 2021). However, Fortin et al. (2021) conducted a recent study and found no evidence of bias in several major journals, including Nature, PNAS, PLoS ONE, New England Journal of Medicine, and Cell. The study found that articles authored by men and women received equal attention, with one exception. The journal Science showed gender bias against women. The median showed 20 more tweets or 3 more news for male first authors compared to female first authors. Nevertheless, this study analyzed a collection of journals that are predominantly focused on multidisciplinary and biomedical research. Therefore, it is recommended to collect additional evidence from diverse fields in order to draw definitive conclusions regarding gender bias in altmetrics.

This work is justified by the two weaknesses previously identified in the literature. On the one hand, the need for more research on the different types of social attention and their meaning. On the other hand, the need for sufficient scientific evidence regarding the possible existence of gender bias in social attention to researchers, especially in relation to the particularities of each scientific field. Therefore, the purpose of our study is to answer the following research question: Is there a gender and/or field bias in the patterns of altmetric mentions -Twitter, news, policy, and Wikipedia mentions- for Spanish researchers with the highest altmetrics?

To answer this question, we examine the most mentioned researchers in Spain, as recorded by the Altmetric platform between 2016 and 2020. Specifically, we compare the top 250 researchers with the highest number of social mentions in the 22 fields of the Web of Science (henceforth WoS) database. Our methodology includes the creation of three rankings based on publication and mention data, as well as nonparametric tests to compare means.

**Theoretical framework**

*Review of the literature on altmetrics*

Most altmetric data improve citations in terms of the accumulation speed after publication (Fang and Costas, 2020). However, except for Mendeley readership, which is moderately correlated with citations (Zahedi and Haustein, 2018), there is a negligible or weak correlation between citations and most altmetrics (Bornmann, 2015; Costas et al., 2015). Therefore, altmetrics may capture other forms of impact rather than citation impact (Wouters et al., 2019).

Altmetrics address the need for researchers to provide evidence of the social impact of their findings. However, it is difficult to measure the social impact of research because there can be a long time between basic research and its practical application (Godin, 2011), and because the obsolescence of results strongly determines the impact of research (Dorta-González and Gómez-Déniz, 2022). Thus, in addition to mentions in social media and mainstream news, altmetrics also include references in public policy documents and recommendations that are more academic than social (Haustein et al., 2016). Furthermore, the variety of indicators and their differences make it advisable to use them separately rather than mixing indicators (Wouters et al., 2019).

Different dimensions of social impact include the following (Thelwall and Nevill, 2018): mentions on Twitter and Facebook may represent social media discussion; blogs and news may reflect attention to newsworthiness; and Wikipedia may account for informational attention. Furthermore, there are different levels of attention depending on the commitment that the social interaction entails (Haustein et al., 2016). For example, retweeting is not the same as writing a post on a blog. Twitter is the most widely used source of altmetric data. There are many studies that provide evidence of its impact, as well as its benefits and limitations. Fang et al. (2022) found that likes were the most common form of engagement, followed by retweets, while quotes and replies were less common. User engagement was associated more with Twitter-based factors, such as the number of users mentioned and followers, than with scholarly factors, such as citations and Mendeley readership. In addition, scholarly tweets in the social sciences and humanities were more likely to generate user engagement than other subject areas. Fang et al. (2020) examined the longevity of Twitter mentions of scientific publications and found that approximately 14.3% of highly tweeted publications were no longer available after 18 months, primarily due to user deletion, suspension, and Twitter account protection. The study highlights the importance of identifying different Twitter dissemination structures when analyzing Twitter metrics for scientific publications.

Some authors have reported a higher presence of altmetrics in the social sciences and humanities than in the natural sciences (Chen et al., 2015). Thus, they have suggested that altmetrics may represent a complement to citations, especially in disciplines of social sciences and humanities. However, research attention and research impact are not synonymous. Social attention is a more complex phenomenon, as it can be motivated by both positive and negative aspects of research (Sugimoto et al., 2017). Other authors have also been interested in the platforms that collect altmetrics, such as Altmetric, Impactstory, and PlumX, mainly in relation to the source of data, the indicators provided, the speed of data accumulation, etc. (Fang and Costas, 2020; Fang et al., 2020; Ortega, 2018). In addition, differences between countries and disciplines according to the coverage of mentions have also been analyzed (Torres-Salinas et al., 2022).

Recently, an important question has been raised about whether altmetric metrics accurately measure the impact or social influence of research. While there is some debate, it is generally accepted that they do not clearly reflect social impact, but rather "unknown impact" (Bornmann et al., 2019) or different types of attention (Thelwall, 2021). Bornmann et al. (2019) investigated the convergent validity of altmetrics by using two REF datasets, publications submitted as research outputs and publications referenced in case studies, to examine the relationship between altmetrics, citation counts, research outputs, and peer REF ratings of research outputs and social impact. While altmetrics have convergent validity for some types of data, such as news media, Facebook, blogs, Wikipedia, and policy-related documents, their validity for capturing social impact, as defined by peer reviewers interested in the causal link between research and action in society, is questionable.

In this regard, Díaz-Faes et al. (2019) introduced the idea of a second generation of social media metrics that characterize the attention of different social media communities to scientific research and their interactions with science-related entities. This approach combines bibliometric and altmetric methods to provide a more comprehensive view of social media research in science, taking into account the interactions and perspectives of social media networks. Recent developments in this area, such as Twitter-based topic-actor networks (Hellsten and Leydesdorff, 2020), highlight a shift towards more interactive and networked perspectives in social media metrics research. These advances highlight the importance of distinguishing primary metrics related to research objects from secondary metrics related to social media entities or users. Further research on altmetrics is needed, especially in light of these recent trends in the field. These new approaches provide a solid framework for potentially categorizing users based on their social media behavior and for benchmarking for comparative analysis among users.

Finally, it should be noted that Gumpenberger et al. (2016) discuss the dangers associated with altmetrics scores and the potential for misuse. The authors highlight the problems associated with composite metrics, particularly in the context of rankings. Publishers and content providers have been quick to adopt the altmetrics scores without criticism, despite the need for proper interpretation. The inclusion of percentiles is a promising approach, but it remains a challenge to differentiate appropriately based on disciplines, document types, and other factors.

*Review of the literature on gender bias*

While gender differences in altmetrics have been observed in research studies, they do not represent an inherent bias in altmetrics themselves. Rather, they reflect broader societal and systemic factors that affect gender representation and visibility in academic contexts. Gender differences in academia, including the underrepresentation of women in most scientific disciplines and their tendency to publish fewer articles over the course of their careers, are well documented but fragmented. Huang et al. (2020) conducted a bibliometric analysis of academic publication careers to provide a comprehensive picture of longitudinal gender differences in performance. They reconstructed the full publication history of over 1.5 million gender-identified authors whose publication careers ended between 1955 and 2010, covering 83 countries and 13 disciplines. Paradoxically, they found that the increase in women's participation in science over the past 60 years has been accompanied by a widening of the gender gap in both productivity and impact. However, they also uncovered two gender invariants: men and women publish at a comparable annual rate and have equivalent career impact for the same body of work. In addition, they showed that differences in the length of publishing careers and dropout rates explain a significant part of the reported career differences in productivity and impact, although productivity differences remain.

Gender equality policies are essential for dismantling the vicious cycle that creates the glass ceiling (Besselaar and Sandström, 2017). These authors suggest that a variety of factors may put female researchers at a disadvantage, reducing their chances of becoming highly productive researchers. Therefore, it is unwise to assume that past trends suggesting that the gender gap in science will gradually narrow over time will hold true. However, Cameron et al. (2016) found no gender differences in research output in the field of ecology after excluding self-citations and non-research active years. This pattern is consistent across disciplines, suggesting that current geographic disparities in research output may be due to rewards for confident behavior and traditional career paths, rather than research impact. Importantly, excluding self-citations and non-research active years would not disadvantage anyone, as self-citations do not necessarily indicate broader impact, and researchers should be judged solely on their research active careers.

Several studies have examined the relationship between gender and altmetrics and found differences in social attention based on gender. There may be differences in social media engagement and dissemination of scholarly work based on gender. For example, research has shown that men tend to have more followers and receive more retweets on Twitter than women (Zhu et al., 2019). Additionally, studies have found that female researchers are more likely to face harassment and online attacks, which may affect their visibility and engagement on social media platforms (Lewis et al., 2020).

A recent study using bibliometrics and altmetrics found that gender differences are still significant (Zhang et al., 2021). The authors state that the differences are small but significant. They distinguish them by the gender of the first author to find the differences, because most of the scientific publications they studied were co-authored by female and male researchers. They found one possible explanation for the gender differences in impact. Male researchers are more likely to value and engage in research aimed primarily at scientific progress, which is more highly cited. Female researchers, on the other hand, are more likely to value and engage in research aimed primarily at contributing to societal progress, which has higher usage.

One large-scale study merits further description. Chapman et al. (2022) conducted a study to examine bias in conservation science journals by comparing the altmetric scores of papers where men and women were first or last authors to a sample of nearly 10,000 papers. Their results showed that, overall, there was no difference in scores between papers authored by men or women. However, the authors noted that high altmetric scores are more likely to impact an academic's career, so they analyzed the scores of papers in the top 10%. They found that papers with male first authors had higher scores than papers with female first authors, while there was no difference in scores for papers with male or female last authors. These findings suggest that gender bias still exists in academia, and that rewarding high altmetric scores may perpetuate this inequality.

Gender differences in altmetrics may also vary by academic discipline. Some studies have found that altmetric attention is distributed differently among male and female researchers depending on the discipline. This suggests that disciplinary norms and the types of research being conducted may influence altmetric results. A paper examining the gender gap in altmetric mentions found that online science dissemination is male-dominated, and that female scientists are less likely to be in the top 25% of the most successful scholars online across the research fields studied (Vásárhelyi et al., 2021). However, a large-scale study of field bias and altmetrics found no bias in major journals, including Nature, PNAS, PLoS ONE, New England Journal of Medicine, and Cell (Fortin et al., 2021). The study found no significant difference in the attention given to articles authored by men and women, except in the case of Science, which showed a gender bias against women in 2018. This bias resulted in male first authors receiving, on average, 88 more tweets or 11 more news mentions than female first authors. The median also showed a gender gap, with male first authors receiving 20 more tweets, or 3 more news mentions than their female counterparts.

Overall, the evidence suggests that there are indeed differences in social attention based on gender, highlighting the need for continued research and efforts to address gender disparities in academia and scholarly communication. In addition, altmetrics may differ depending on the characteristics and dynamics of each field, as some fields may tend to have a more active presence on social media platforms and receive higher levels of social attention.

**Materials and Methods**

*Data*

We have analyzed the researchers in Spain with the most mentions on the Altmetric platform between 2016 and 2020. Specifically, we considered the top 250 researchers in each of the 22 fields of the WoS. The top 250 most socially influential researchers in each field (hereafter Top Influential Researchers) were selected based on the Altmetric Attention Score (AAS). The AAS is an aggregated and weighted metric that reflects the social impact of research output, including mentions in news outlets, social media platforms, blogs, and other online sources.

The data source is the open dataset of Arroyo-Machado et al. (2022). This dataset includes the publications with any altmetric authored by researchers affiliated to Spanish institutions, along with their associated altmetric mentions. In a first file, each record represents a scientific publication (article, review, or letter) from the WoS database, published between 2016 and 2020. It provides bibliographic metadata, its subject field, and a list of altmetric indicators from the Altmetric platform. The second file contains the 250 Top Influential Researchers in the 22 fields of the WoS. Each record contains the disambiguated author's full name, ORCID identifier, affiliation, and a list of publications that link to the first file. The gender of each author was inferred from the second file through a manual process, relying primarily on the researchers' names to infer their gender. We also used additional information, such as ORCID identifiers and affiliation homepages, to gather more details about the researchers in cases of doubt.

In addition, some considerations can be made. Since the scientific fields in the WoS differ in the number of socially influential authors, it would be preferable to use the authors in the highest percentiles according to their AAS. However, the available information includes the same number of researchers for each field without providing the total number of authors in the field, so we were limited to working with the same sample size in each field (i.e., 250).

Some researchers are listed in more than one research field, so instead of a total of 5500 researchers (i.e., 250 Top Influential Researchers in 22 fields) there are 4195 different researchers in our dataset. Catalonia and the Community of Madrid have 56% of the total number of these researchers, Andalusia, the Valencian Community, and the Basque Country have 25% of the total number, and the rest is distributed among the other regions. This distribution is obviously determined by the size of the region and the number of universities and research centers. In fact, a total of six universities and research centers account for 35% of the total number of researchers with the greatest social impact (CSIC and the Universities of Barcelona, Autónoma de Barcelona, Complutense de Madrid, Pompeu Fabra, and Autónoma de Madrid, in that order). To contextualize the information we have in the dataset for each WoS field, the gender distribution of these researchers is shown in Table 1 for each field. It is noteworthy that there is an unequal gender distribution among the most influential researchers, as shown in Table 1.

*Table 1. Number of male and female Top Influential Researchers, across fields*

| Web of Science fields | Female | | Male | |
|---|---|---|---|---|
| *Agricultural Sciences* | *129* | *52%* | *121* | *48%* |
| Arts & Humanities | 83 | 33% | 167 | 67% |
| Biology & Biochemistry | 66 | 26% | 184 | 74% |
| Chemistry | 57 | 23% | 193 | 77% |
| Clinical Medicine | 71 | 28% | 179 | 72% |
| Computer Science | 39 | 16% | 211 | 84% |
| Economics & Business | 66 | 26% | 184 | 74% |
| Engineering | 53 | 21% | 197 | 79% |
| Environment/Ecology | 78 | 31% | 172 | 69% |
| Geosciences | 53 | 21% | 197 | 79% |
| Immunology | 101 | 40% | 149 | 60% |
| Materials Science | 71 | 28% | 179 | 72% |
| Mathematics | 44 | 18% | 206 | 82% |
| Microbiology | 83 | 33% | 167 | 67% |
| Molecular Biology & Genetics | 73 | 29% | 177 | 71% |
| Neuroscience & Behavior | 84 | 34% | 166 | 66% |
| Pharmacology & Toxicology | 112 | 45% | 138 | 55% |
| Physics | 46 | 18% | 204 | 82% |
| Plant & Animal Science | 61 | 24% | 189 | 76% |
| Psychiatry/Psychology | 107 | 43% | 143 | 57% |

| | | | | | |
|---|---|---|---|---|---|
| Social Sciences, General | | 92 | 37% | 158 | 63% |
| Space Sciences | | 54 | 22% | 196 | 78% |
| | | 1,623 | 30% | 3,877 | 70% |

*Note: In bold italics is the only field of the WoS with a higher presence of female than male Top Influential Researchers. Shaded rows are those fields with a higher proportion of women than average.*

*Methodology*

To identify potential gender disparities, we compared the distribution of publications within the WoS with mentions across the different altmetrics. We examined publications with mentions in all four altmetrics, stratifying the analysis by gender and field. In addition, we examined how the number of publications corresponded to the number of mentions in each altmetric, differentiating by gender.

The mentions that publications receive on different digital platforms make it possible to quantify and measure the impact of scientific research. In addition, they have different dimensions and meanings depending on the characteristics of the platforms used for their distribution and the type of audience that accesses the information. As a result, each type of digital platform where a scientific article is referenced reflects a different type of influence that the academic one, making it possible to quantify the task of disseminating knowledge to society in several of its dimensions.

Arroyo-Machado et al. (2022) identified four dimensions of social influence that can be traced in different platforms, namely, media, political, social, and educational. These types of influence are described in Table 2. Note that these are not perfectly delimited influences, but rather a simplified approximation of a complex reality. In some cases, there may be overlap between dimensions. The development and use of these indicators is essential for evaluating research and determining the extent and impact of research on societal stakeholders.

*Table 2. Dimensions in the influence of scientific research*

| Dimension | Proxy | Description |
|---|---|---|
| Media influence | News | The media play an important role in communicating science to a non-specialized audience. One way to quantify media attention is to determine the number of mentions of scientific publications receive in the main digital newspapers. |
| Political influence | Policy | This is the impact on public policy makers. The number of references or citations in public reports may reflect the interest in public policy that it generates. |
| Social influence | Twitter | This is the influence exerted on a public that is not specialized in science and that encompasses society in a cross-cutting way. One of the most popular and global social media is Twitter, a digital reflection of a part of society. The mentions or the times a scientific publication is shared on this social media may reflect the public interest or the debate it sparks. |
| Educational influence | Wikipedia | The use of scientific results in education is an expression of the impact of research results on society. Currently the most relevant platform with an encyclopedic approach is Wikipedia. The number of mentions of publications in Wikipedia entries may reflect the ability to influence educational contexts. |

The second column of the table contains variable proxies that can serve as measures of the concepts in the first column. For example, media influence can be measured by media reputation or audience coverage, but these data can be difficult to obtain. Therefore, a simpler proxy variable (News) based solely on the number of media news items was used to reflect the reality being measured. However, this approach ignores factors such as the audience size. Similarly, political influence is approximated by the number of mentions in public policy documents, but this approach ignores important aspects such as the impact of the policy and the organization promoting it. For social media, Twitter has been used as a proxy to measure the social influence of scientific research conversations, but there are other social networks and factors, such as the sentiment analysis and the number of followers, that are not considered. Similarly, the educational impact of scientific research is approximated using Wikipedia, which is widely used but not the only source of information.

Since the number of mentions in each dimension and field of research is so heterogeneous -i.e. the number of mentions in News is not comparable to the number of mentions in Policy- it is necessary to look for measures that allow relative comparisons. For this reason, we decided to assign a ranking position to each of the 5500 researchers according to their mentions in each dimension. In this way, we can subsequently analyze the distribution of these rank positions across scientific fields, distinguishing those fields with researchers with the most mentions from those with the fewest.

We propose the following three rankings to classify researchers according to their social orientation and its magnitude.

a) *Social attention orientation (Ranking #1)*. Based on the proportion of publications with social mentions (i.e., number of publications with social mentions in a given altmetric / total number of publications with mentions in any of the four studied altmetrics). It is a measure in the interval [0, 1]. The lower bound (0) means that none of the researcher's publications have mentions in that altmetric, while the upper bound (1) means that all of the researcher's publications have mentions in that altmetric. Thus, by ordering the researchers in a ranking based on these data, we measure how social attention-oriented each researcher is.

b) *Level of social attention (Ranking #2)*. Based on mentions per publication (i.e., number of social mentions in a given altmetric / total number of publications with mentions in any of the four studied altmetrics). This is a measure in the interval [0, +∞). The lower bound (0) means that none of the researcher's publications have any social mentions in this altmetric, while the upper bound (+∞) means that the set of publications is very highly mentioned in this altmetric. Thus, by ordering the researchers in a ranking based on these data, we measure the level of social attention of each researcher.

c) *Intensity of social attention (Ranking #3)*. Based on mentions per mentioned publication (i.e., number of social mentions in a given altmetric / total number of publications with social mentions in that altmetric). This is a combination of the previous two measures. Researchers whose publications have no social mentions in a given altmetric are, by definition, outside this measure, which is defined in the interval [1, +∞). The lower bound (1) means that each mentioned publication has received one social mention, while the upper bound (+∞) means that the set of mentioned publications is also very highly mentioned. Therefore, by ordering researchers in a ranking based on these data, we measure the degree of intensity in the social attention of each researcher. This ranking will be closer to ranking #2 the better the researchers are positioned in ranking #1 (ratio close to one). However, it will be further away from ranking #2 the worse the researchers are positioned in ranking #1 (ratio close to zero). In addition, this ranking #3 favors researchers with a reduced number of highly mentioned publications in a particular altmetric.

Next, to verify that these three rankings measure different aspects of social attention, we calculated the degree of relationship between them through Spearman's rank correlation coefficient with the information collected through the four sources of altmetrics.

The position of each researcher, according to the three proposed rankings, in each of the four dimensions was determined using the Microsoft Excel function 'RANK.EQ'. Graphical analyses of the distribution of these rankings, non-parametric tests of equality of distributions (Kruskal-Wallis tests), Spearman's contrasts, and non-parametric mean-comparison-tests by gender (Mann-Whitney U tests) were performed using Stata 17.

**Results**

*Field differences on social impact of research*

Figures 1 to 3 show the distribution of the positions in the global ranking (#1, #2 or #3) of the 250 Top Influential Researchers in each field for each of the four sources (news, policy, Twitter, and Wikipedia). Those fields with a higher proportion of researchers in the top positions have more social attention according to this ranking, while those with a higher proportion in the bottom positions have less social attention. Note that, in the rankings #1 and #2 there are always 5500 observations, but many that share the last position in the ranking because they have not received any mention. Therefore, it seems that some rankings have fewer observations and there are scale changes in the figures. In the case of ranking #3, on the other hand, all researchers who did not receive mentions are excluded, so the number of observations is reduced to 4853 for news, 1083 for policy, and 1765 for Wikipedia. In the case of Twitter, on the other hand, since the 250 Top Influential Researchers in each field received some mentions, the sample remains complete with the 5500 researchers in ranking #3.

Figure 1 shows the social attention orientation of the Top Influential Researchers by discipline in each dimension compared to the other researchers, using the box-and-whisker plot. What would be expected if there were no differences by fields is a uniform distribution. Instead, the farther to the left of the boxes, the better positioned in the ranking, and the farther to the right, the worse positioned. Hence, distributions that are skewed to the left show better positions in the ranking, and thus a greater social orientation of their top researchers.

*Figure 1. Distribution of ranking #1 for News, Twitter, Wikipedia, and Policy, by field. Ranking based on the proportion of publications with social mentions (social attention orientation)*

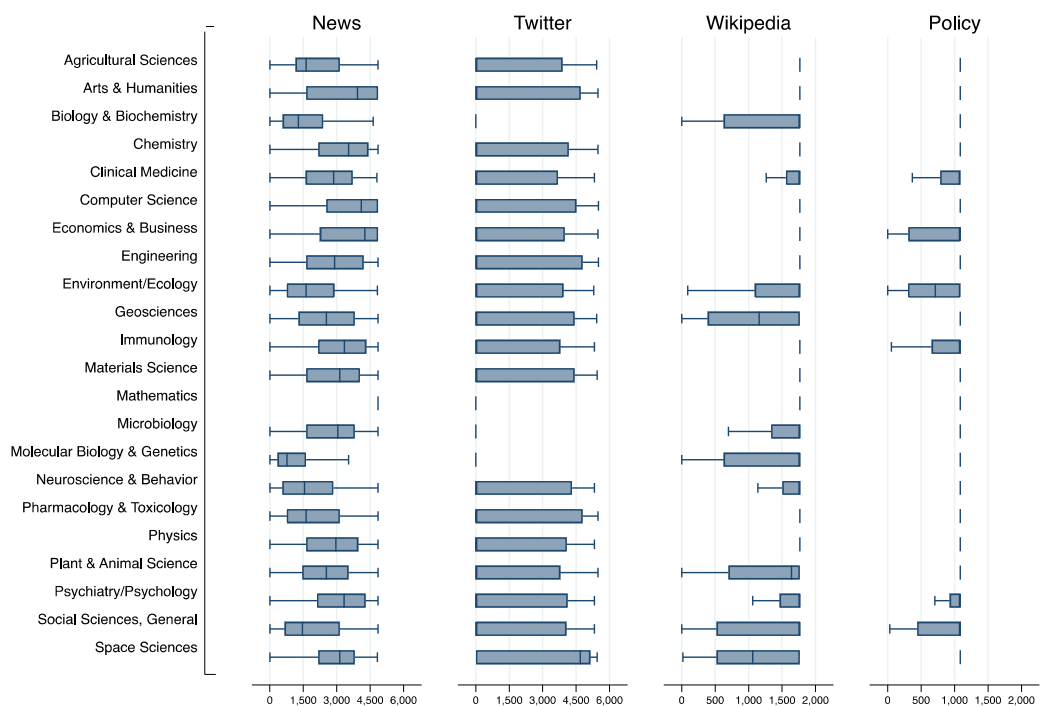

Important differences between fields are clearly shown in Figure 1. For example, the fields that receive the most media attention in the news are those related to biology (Molecular Biology & Genetics, Biology & Biochemistry). However, the fields that receive the least media attention in the news are, in order, Mathematics, Computer Science, Economics & Business, and Arts & Humanities. Furthermore, the most discussed fields on Twitter are also those related to biology (Biology & Biochemistry, Microbiology, Molecular Biology & Genetics).

Similarly, Figure 2 shows the amount of social attention given to the Top Influential Researchers by discipline in each dimension compared to the other researchers, using the box-and-whisker plot. Again, what would be expected if there were no differences by field is a uniform distribution. Instead, distributions that are skewed to the left show better positions in the ranking and, therefore, greater social attention for their top researchers.

*Figure 2. Distribution of ranking #2 for News, Twitter, Wikipedia, and Policy, by field. Ranking based on mentions per publication (level of social attention)*

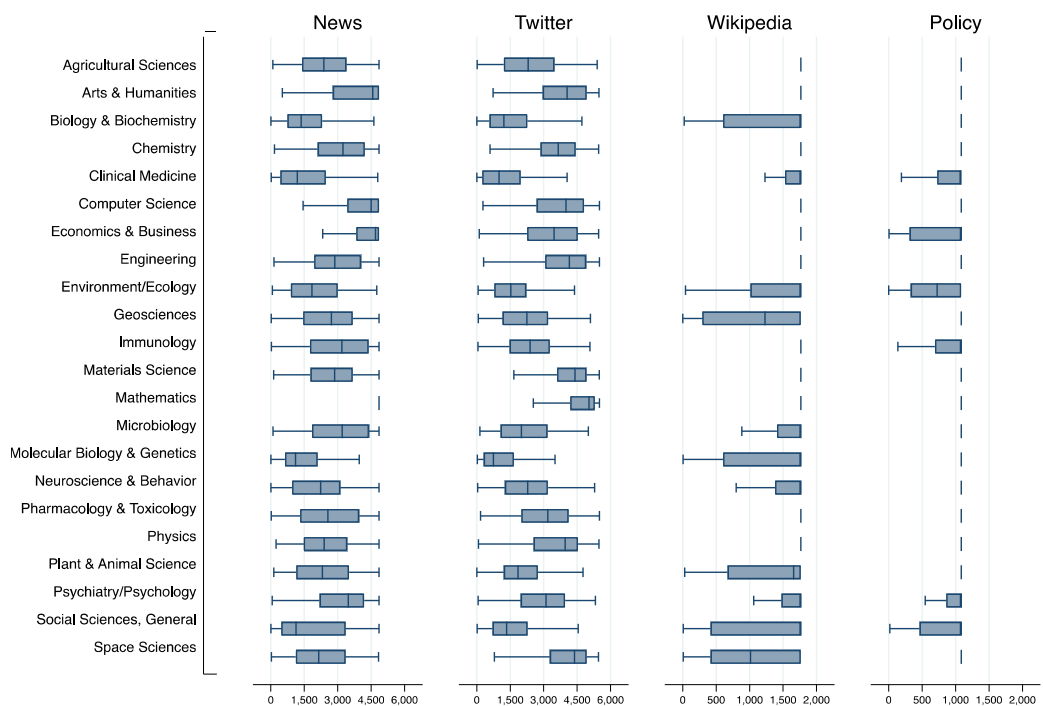

Important differences between fields are again clearly shown in Figure 2. For example, the fields that receive the most media coverage in the news are again those related to biology (Molecular Biology & Genetics, Biology & Biochemistry), together with Clinical Medicine. However, the fields that receive the least media coverage in the news are, in this order, Mathematics, Economics & Business, Computer Science, and Arts & Humanities. In addition, the most popular fields on Twitter are Molecular Biology & Genetics and Clinical Medicine. The least popular fields on Twitter are Mathematics and Materials Science.

Figure 3 shows the degree of intensity in the social attention of the Top Influential Researchers by discipline in each dimension compared to the other researchers, using the box-and-whisker plot. Again, a uniform distribution would be expected if there were no differences by fields. The further to the left of the boxes, the higher the position in the ranking and, therefore, the greater the intensity of social attention of the researchers in that discipline.

*Figure 3. Distribution of ranking #3 for News, Twitter, Wikipedia, and Policy, by field. Ranking based on mentions per mentioned publication (intensity in the social attention)*

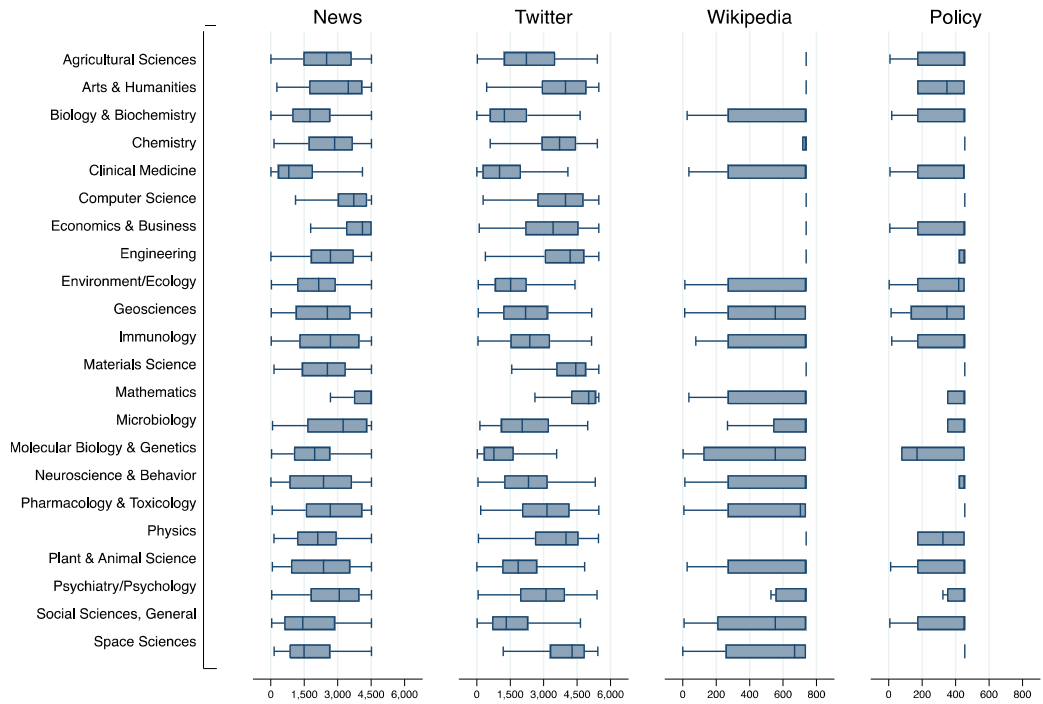

Again, Figure 3 clearly shows important differences between fields. For example, the fields that receive the most media attention in the news, as the publications are mentioned many times, are Clinical Medicine and Social Sciences. However, the fields that receive the least media attention in the news are, in this order, Mathematics, Economics & Business, and Computer Science. Furthermore, the most popular fields on Twitter, with many more tweets and retweets than the others, are Molecular Biology & Genetics, Clinical Medicine, Biology & Biochemistry, and Social Sciences. The least popular fields on Twitter are again Mathematics and Materials Science.

For each of the three proposed measures and the four dimensions analyzed, we verified that the graphical differences observed in the box plots between fields are statistically significant using nonparametric Kruskal-Wallis tests, as shown in Table 3.

*Table 3. Kruskal-Wallis tests for equality of distribution among WoS fields*

|  | News | | Twitter | | Wikipedia | | Policy | |
| --- | --- | --- | --- | --- | --- | --- | --- | --- |
|  | Chi2= 21 | Sig. | Chi2= 21 | Sig. | Chi2= 21 | Sig. | Chi2= 21 | Sig. |
| Ranking #1 | 1300.64 | **** | 590.07 | **** | 326.48 | **** | 514.01 | **** |
| Ranking #2 | 1767.57 | **** | 593.34 | **** | 2241.32 | **** | 528.57 | **** |
| Ranking #3 | 771.74 | **** | 52.53 | **** | 2233.50 | **** | 139.99 | **** |

*Note:* **** *The equal distribution hypothesis among WoS fields can be rejected at any level below 0.1%.*

*Independence of dimensions and measures (correlations between rankings)*

It could be argued that those who receive the most social attention are the same researchers who publish the most. To verify this statement we calculated Spearman's rank correlation coefficients between the three proposed rankings and the one obtained by ranking researchers according to their scientific productivity in the WoS with some altmetric mention, without differentiating fields of research (see column 'Ranking by Num Pub' in Table 4). The values of these coefficients show that there is little correlation between the ranking by publications and the three rankings by attention in each of the four dimensions.

To check whether the three proposed altmetric rankings measure the same aspect of social attention, we also computed Spearman's rank correlation coefficients between all of them in the four dimensions (Table 4 and Supplementary Table 1). In this way, we checked that the correlation between almost all of them is very low (see Supplementary Table 1), indicating that they measure different aspects of social attention. There is only an almost perfect correlation between rankings #1 and #2 for policy and Wikipedia, and between rankings #2 and #3 for news and Twitter (see Table 4). However, there are very few mentioned publications and mentions on policy documents and Wikipedia pages, so it is not surprising that both rankings are closely related. In any case, we can assume that these rankings generally measure different aspects of social attention.

*Table 4. Spearman's correlation coefficients between the different rankings*

| **News** | Ranking by Num Pub | Ranking #1 | Ranking #2 | Ranking #3 | **Policy** | Ranking by Num Pub | Ranking #1 | Ranking #2 | Ranking #3 |
|---|---|---|---|---|---|---|---|---|---|
| Ranking #1 | -0.26 | 1 | | | Ranking #1 | 0.17 | 1 | | |
| Ranking #2 | -0.23 | 0.73 | 1 | | Ranking #2 | 0.18 | 0.99 | 1 | |
| Ranking #3 | -0.27 | 0.17 | 0.87 | 1 | Ranking #3 | -0.04* | 0.22 | 0.58 | 1 |
| **Twitter** | Ranking by Num Pub | Ranking #1 | Ranking #2 | Ranking #3 | **Wikipedia** | Ranking by Num Pub | Ranking #1 | Ranking #2 | Ranking #3 |
| Ranking #1 | -0.27 | 1 | | | Ranking #1 | 0.12 | 1 | | |
| Ranking #2 | -0.12 | 0.36 | 1 | | Ranking #2 | 0.13 | 0.99 | 1 | |
| Ranking #3 | -0.20 | 0.31 | 0.99 | 1 | Ranking #3 | -0.11 | 0.32 | 0.61 | 1 |

*Note: Statistically significant at 0.1% except for the coefficient marked with ***

*Gender differences on social attention of researchers*

To identify potential gender differences, we compared the distributions of publications in WoS with mentions in any of the altmetrics and the publications with some mentions in the four altmetrics analyzed for men and women and by scientific field. We also compared the distributions of the number of publications with the number of mentions in each altmetric (see Supplementary Figure 1 and Supplementary Figure 2). These figures show that women systematically have fewer publications and mentions than men, except in the fields of Psychiatry/Psychology and Social Sciences, where the pattern is reversed. However, these differences may not be statistically significant at the mean level.

In order to analyze whether there are statistically significant mean differences between men and women at different levels of significance, given that the variables are not normally distributed, non-parametric mean comparison tests (Mann-Whitney U tests) are performed to see whether the apparent differences by gender in Supplementary Figure 1 and Supplementary Figure 2 are indeed statistically significant. Many of them are not, as can be seen in Table 5 and able 6.

*Table 5. Mann-Whitney U tests by gender for the number of publications with mentions in any of the altmetrics, and the number of publications with some mention in News, Twitter, Wikipedia, and Policy, by field*

|  | Num Pub | | News | | Twitter | | Wikipedia | | Policy | |
| --- | --- | --- | --- | --- | --- | --- | --- | --- | --- | --- |
|  | z | Sig. | z | Sig. | z | Sig. | z | Sig. | z | Sig. |
| Agricultural Sciences | -2.26 | ** | -0.80 | | -2.37 | ** | -0.31 | | -0.28 | |
| Arts & Humanities | -0.99 | | 0.33 | | -1.36 | | 0.75 | | -1.23 | |
| Biology & Biochemistry | -3.92 | *** | -3.14 | *** | -3.77 | *** | -2.46 | ** | -0.74 | |
| Chemistry | -3.80 | *** | -1.14 | | -3.88 | *** | 0.04 | | 0.92 | |
| Clinical Medicine | -3.63 | *** | -2.07 | ** | -3.62 | *** | -2.54 | ** | -1.23 | |
| Computer Science | -2.03 | ** | 0.40 | | -2.06 | ** | -1.18 | | -1.52 | |
| Economics & Business | -2.79 | *** | 0.10 | | -2.40 | ** | -1.22 | | 0.47 | |
| Engineering | -1.14 | | 0.28 | | -1.31 | | -0.11 | | -0.71 | |
| Environment/Ecology | -0.16 | | 0.51 | | -1.67 | * | -3.20 | *** | 0.98 | |
| Geosciences | -2.47 | *** | -1.53 | | -2.48 | ** | -2.53 | ** | 0.41 | |
| Immunology | -4.08 | *** | -0.14 | | -4.11 | *** | -1.07 | | 0.08 | |
| Materials Science | -2.53 | ** | -3.00 | *** | -2.72 | *** | -1.84 | * | 1.19 | |
| Mathematics | 0.22 | | 0.60 | | -0.11 | | 0.05 | | 0.78 | |
| Microbiology | -2.71 | *** | -2.10 | ** | -2.71 | *** | -1.85 | * | -0.54 | |
| Molecular Biology & Genetics | -4.50 | *** | -2.94 | *** | -4.43 | *** | -2.76 | *** | -0.88 | |
| Neuroscience & Behavior | -0.43 | *** | -2.96 | *** | -4.31 | *** | 0.57 | | -1.98 | ** |
| Pharmacology & Toxicology | -4.53 | *** | -1.50 | | -4.74 | *** | -0.87 | | 0.21 | |
| Physics | -3.01 | *** | -1.75 | * | -3.04 | *** | -2.43 | ** | -0.83 | |
| Plant & Animal Science | -3.08 | *** | -1.99 | ** | -3.11 | *** | -2.31 | ** | 0.39 | |
| Psychiatry/Psychology | 0.36 | | -0.22 | | 0.33 | | 0.06 | | 1.96 | * |
| Social Sciences, General | -0.10 | | 2.44 | ** | 0.04 | | -1.50 | | 2.09 | ** |
| Space Sciences | -1.65 | * | -1.41 | | -1.39 | | -1.41 | | -0.53 | |

*Note: Statistically significant differences in means between men and women at the 1% ( \*\*\* ); 5% ( \*\* ); or 10% ( \* ) significance level. A negative sign indicates that men are more likely than women to have publications with mentions.*

Regarding the number of publications with any altmetric mention, significant gender differences were found in 16 out of the 22 fields analyzed (Table 5). Moreover, the most egalitarian dimension of social attention, where there is a larger number of fields without gender differences, is the public policy, where significant differences were found in only 3 fields, followed by the mainstream news and Wikipedia, where significant gender differences were found in 9 out of the 22 fields. On the contrary, Twitter is the least egalitarian dimension of social attention, with a smaller number of fields without gender differences, specifically 6 out of the 22 analyzed. From the sign of the significant statistics, it can be seen that: a) the only field in mainstream news where women are mentioned more than men is Social Sciences; b) in policy, women have more mentions than men in Psychiatry/Psychology and Social Sciences; c) otherwise, men consistently have more publications and mentions than women.

*Table 6. Mann-Whitney U tests by gender for number of mentions in News, Twitter, Wikipedia, and Policy, by field*

|  | News mentions | | Twitter mentions | | Wikipedia mentions | | Policy mentions | |
| --- | --- | --- | --- | --- | --- | --- | --- | --- |
|  | z | Sig. | z | Sig. | z | Sig. | z | Sig. |
| Agricultural Sciences | 0.71 |  | -0.77 |  | -0.28 |  | -0.11 |  |
| Arts & Humanities | 0.21 |  | -1.13 |  | 0.84 |  | -1.23 |  |
| Biology & Biochemistry | 0.17 |  | 0.04 |  | -2.09 | ** | -0.75 |  |
| Chemistry | 1.64 |  | -4.55 | *** | 0.02 |  | 0.92 |  |
| Clinical Medicine | 3.01 | *** | -3.95 | *** | -2.70 | *** | -1.28 |  |
| Computer Science | -0.52 |  | -0.60 |  | -1.22 |  | -1.52 |  |
| Economics & Business | -0.11 |  | 0.13 |  | -1.23 |  | 0.64 |  |
| Engineering | -0.24 |  | -1.21 |  | -0.17 |  | -0.73 |  |
| Environment/Ecology | 0.87 |  | -2.77 | *** | -2.89 | *** | 1.09 |  |
| Geosciences | 0.14 |  | -1.36 |  | -2.74 | *** | 0.53 |  |
| Immunology | 1.12 |  | -2.46 | ** | -1.00 |  | 0.26 |  |
| Materials Science | -1.42 |  | -0.79 |  | -1.87 | * | 1.19 |  |
| Mathematics | 0.98 |  | -2.17 | ** | 0.06 |  | 0.77 |  |
| Microbiology | 0.89 |  | -0.83 |  | -1.54 |  | -0.59 |  |
| Molecular Biology & Genetics | 0.77 |  | -1.07 |  | -1.92 | * | -0.88 |  |
| Neuroscience & Behavior | 0.32 |  | -1.67 | * | 0.44 |  | -2.00 | ** |
| Pharmacology & Toxicology | 0.95 |  | -2.49 | ** | -0.77 |  | 0.28 |  |
| Physics | 0.24 |  | -1.07 |  | -2.48 | ** | -0.83 |  |
| Plant & Animal Science | 0.30 |  | -2.30 | ** | -2.15 | ** | 0.31 |  |
| Psychiatry/Psychology | -0.59 |  | -1.26 |  | 0.23 |  | 1.93 | * |
| Social Sciences, General | 3.74 | *** | -3.96 | *** | -1.75 | * | 2.03 | ** |
| Space Sciences | -1.35 |  | -1.41 |  | -1.57 |  | -0.53 |  |

*Note: Statistically significant differences in means between men and women at the 1% (***); 5% (**); or 10% (*) significance level. A negative sign indicates that men are mentioned more often than women*

In terms of the total number of mentions (able 6), the most egalitarian dimensions are the mainstream news and policy. In the former, there are significant differences in only two fields, where women are also the most mentioned, and in the latter in three fields, where men are mentioned more than women in only one field. In contrast, Twitter and Wikipedia mentions show gender differences in 9 of the 22 fields, with men consistently receiving more mentions than women.

However, it is important to note that the data source contains numerous instances of zero mentions in both Wikipedia and policy, as shown in Figure 4 and Figure 5, especially for women in the case of Wikipedia and for men in the case of policy. Therefore, looking only at mentions in news and Twitter (columns 2 and 3 in Table 5; and columns 1 and 2 in able 6) provides a more accurate understanding of gender bias in social attention. Thus, out of the 44 scenarios considered, including social attention on news and Twitter (22 x 2), gender differences were observed in 57% (25) of the cases related to publications with mentions and in 25% (11) of the cases related to total mentions. Therefore, we conclude that, for most fields and dimensions with enough non-zero data, gender differences in social attention are not predominant, but still present. More specifically, out of the 88 cases analyzed for the news and Twitter altmetrics, 36 cases show gender differences (which is 41% of the cases).

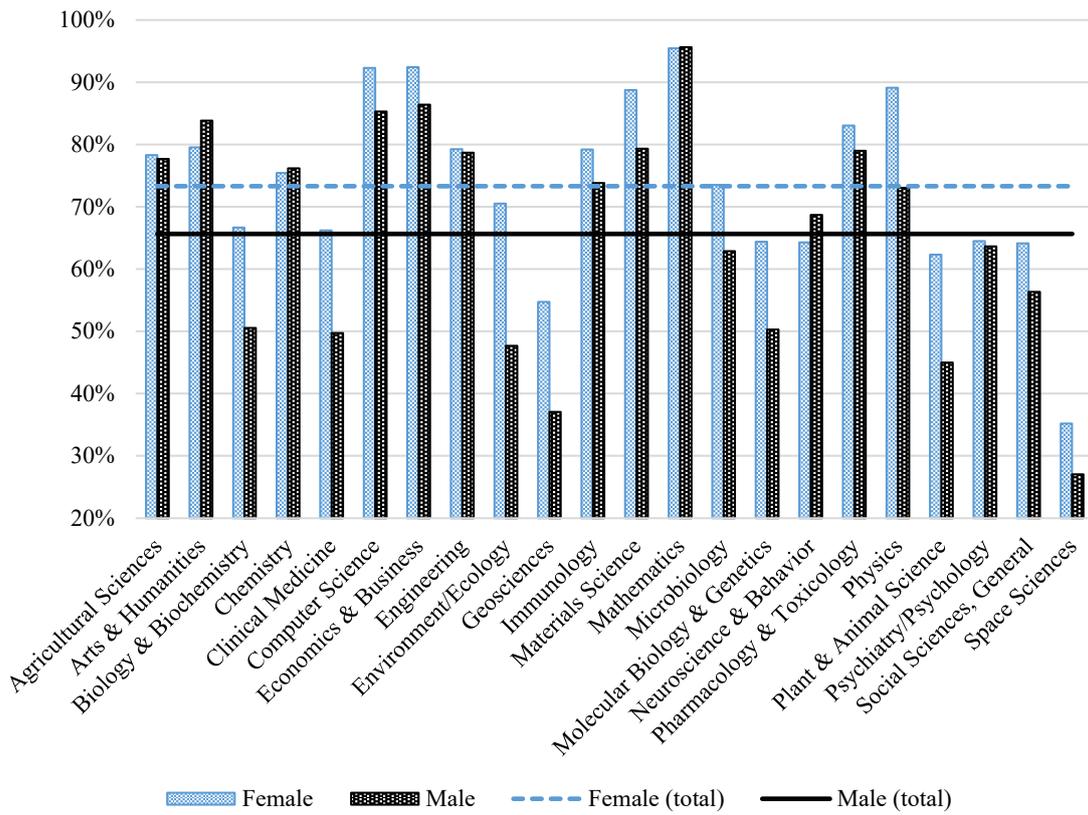

*Figure 4. Proportion of researchers with zero mentions in Wikipedia, by gender*

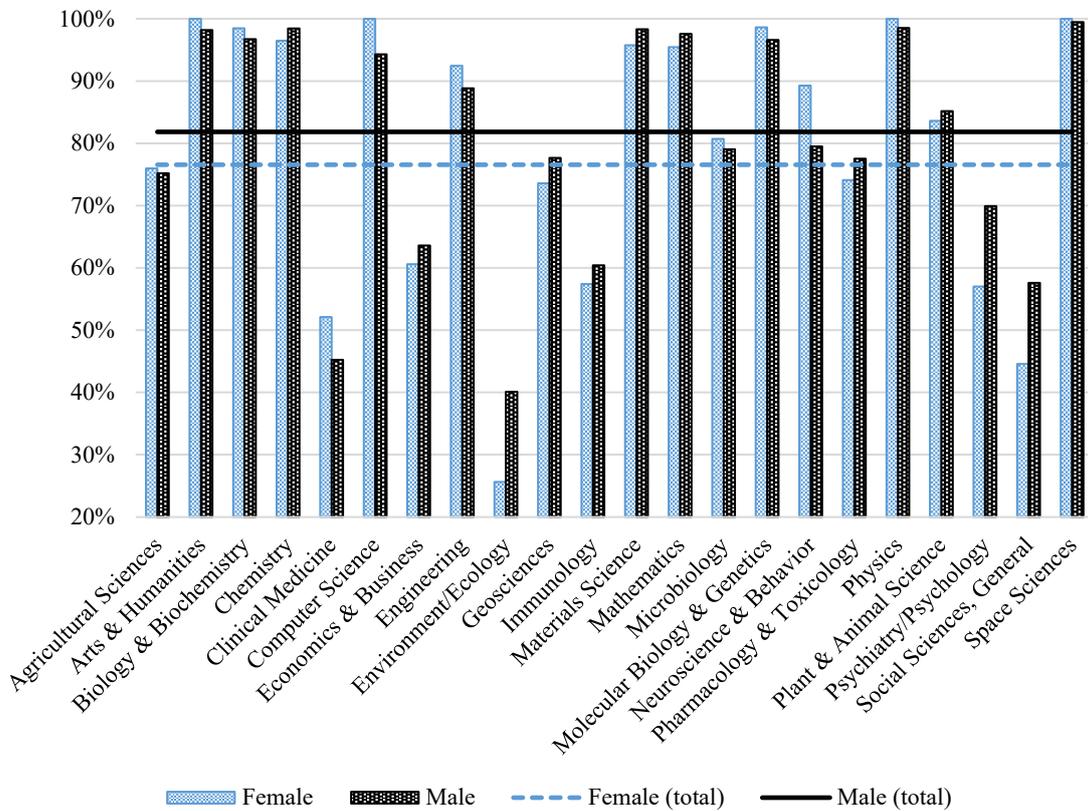

*Figure 5. Proportion of researchers with zero mentions in Policy, by gender*

When Mann-Whitney U tests by gender are performed again to analyze the other two dimensions, i.e. Wikipedia and policy excluding researchers with zero mentions, there is less evidence of gender differences in social attention (Table 7). For publications mentioned in Wikipedia, significant differences are found in only 3 of the 22 research fields, and for mentions, significant differences are found in 4 fields. In the case of publications mentioned in policy documents, significant differences are found in 3 fields, and in terms of mentions, significant differences are found in 2 fields. Thus, out of the 88 cases analyzed for the Wikipedia and policy altmetrics, 12 cases show gender differences (i.e. 14% of the cases).

*Table 7. Mann-Whitney U test by gender, excluding researchers with zero mentions, for number of publications and mentions in Wikipedia and Policy, by field*

|  | Wikipedia publications | | Wikipedia mentions | | Policy publications | | Policy mentions | |
| --- | --- | --- | --- | --- | --- | --- | --- | --- |
|  | z | Sig. | z | Sig. | z | Sig. | z | Sig. |
| Agricultural Sciences | -1.80 | * | -1.28 |  | -1.07 |  | 0.18 |  |
| Arts & Humanities | -1.09 |  | 0.03 |  | . |  | . |  |
| Biology & Biochemistry | -1.14 |  | 0.15 |  | . |  | -0.84 |  |
| Chemistry | -0.59 |  | -0.68 |  | . |  | . |  |
| Clinical Medicine | -1.11 |  | -1.63 |  | -0.86 |  | -0.99 |  |
| Computer Science | -0.45 |  | -0.98 |  | 0.24 |  | . |  |
| Economics & Business | 1.79 | * | 1.24 |  | 0.24 |  | 0.88 |  |
| Engineering | -0.21 |  | -0.72 |  | 1.14 |  | 0.77 |  |
| Environment/Ecology | 0.08 |  | 1.07 |  | -1.72 | * | -1.46 |  |
| Geosciences | -1.14 |  | -1.65 | * | -1.36 |  | -0.57 |  |
| Immunology | -0.83 |  | -0.26 |  | -1.36 |  | -0.67 |  |
| Materials Science | -1.53 |  | -1.83 | * | . |  | . |  |
| Mathematics | . |  | 0.45 |  | 0.42 |  | -0.22 |  |
| Microbiology | -1.21 |  | 0.47 |  | -2.16 | ** | -2.39 | ** |
| Molecular Biology & Genetics | -3.02 | *** | 0.00 |  | . |  | -0.27 |  |
| Neuroscience & Behavior | -0.57 |  | -1.09 |  | -0.79 |  | -1.02 |  |
| Pharmacology & Toxicology | -0.59 |  | 0.28 |  | -3.26 | *** | -2.52 | ** |
| Physics | -1.56 |  | -1.97 | ** | . |  | . |  |
| Plant & Animal Science | -0.36 |  | 0.09 |  | 1.06 |  | 0.09 |  |
| Psychiatry/Psychology | 0.95 |  | 1.58 |  | -0.50 |  | -0.54 |  |
| Social Sciences, General | -1.25 |  | -2.12 | ** | 0.62 |  | 0.47 |  |
| Space Sciences | -0.81 |  | -1.10 |  | . |  | . |  |

*Note: Statistically significant differences in means between men and women at the 1% ( \*\*\* ); 5% ( \*\* ); or 10% ( \* ) significance level. A negative sign indicates that men are mentioned more often than women.*

**Discussion and conclusions**

The aim of this work was to identify differences between disciplines and by gender in the public attention given to researchers. For this purpose, the most influential researchers in Spain were analyzed. Four dimensions of social influence and three measures of social attention were used, and significant differences between disciplines were observed. Evidence was obtained that the sources of mentions analyzed characterize different and independent aspects of social influence. Finally, the four dimensions (media influence, political influence, social media influence, and educational influence) were used to investigate possible gender biases in social influence.

We found important differences across fields and dimensions. Altmetrics can vary depending on the characteristics and dynamics of each field. Some fields tend to have a more active presence on social media platforms and may receive higher levels of social attention (Vásárhelyi et al., 2021). This may be due to factors such as the accessibility of research findings, the presence of a broader audience interested in the field, or the ease of communicating research through practical applications (Zhang et al., 2021). In contrast, fields with more specialized or niche topics may have a smaller audience or a different communication culture, resulting in comparatively lower altmetrics scores. Furthermore, disciplines with a longer publication cycle or a greater reliance on traditional citation-based metrics may have lower levels of social attention.

In terms of the number of articles with any mention, gender differences predominate, with men being more likely than women to have publications with mentions in 16 of the 22 fields examined (73%). However, the percentage drops to 41% when mentions in the two main sources of social attention (news and Twitter) are considered. Among the dimensions of social influence, media attention (news) exhibited the highest level of gender neutrality, whereas social media attention (Twitter mentions) demonstrated the least gender neutrality. Twitter is a social media platform that has the potential to help scientists disseminate their research for social impact (Kapp et al., 2015). Indeed, all of the most influential Spanish researchers in the sample have some mentions on this platform. However, we found evidence that it is also the least egalitarian dimension of social attention that researchers receive, as gender differences are statistically significant in most research fields.

In some disciplines, it is common for research to be mentioned in public policy reports. This typology of papers with potential application to public policy is characterized by a high immediacy of application to social problems, rapid incorporation into knowledge and rapid aging (Rowlands, 2009). However, it was noted that the data used in our study has a significant number of zero values, which limits the ability to draw conclusive results about the true extent of gender bias in public policy reports.

These findings are relevant for policymakers and science managers, as identifying research with the greatest social impact allows research funding strategies and programs to be designed in a more egalitarian way, especially from a gender perspective (Besselaar and Sandström, 2017). Furthermore, studies on gender and altmetrics can help to identify potential gender biases in the dissemination and visibility of scholarly work. Research has shown that female-authored articles tend to receive lower altmetric scores than male-authored articles in some journals (Fortin et al., 2021), which may reflect gender biases in the way scholarly work is promoted and shared online. However, our results show that such gender differences are in the minority, at least as far as the media and social networks are concerned.

The new framework for evaluating scientific publications is the result of a combination of traditional bibliometric and altmetric measures. These measures have been supported by the European Commission (Wilsdon et al., 2017). In this report, it is recommended that evaluators incorporate these indicators of social attention, while being aware of their limitations, in order to provide a more comprehensive quantitative measure of the research. Altmetrics are also of particular interest for the evaluation of Open Science platforms in the new scientific contexts covered by the concept of Responsible Research and Innovation (RRI). The RRI concept in the Horizon 2020 'Science with and for Society Program' (Delaney and Tornasi, 2020) aims to reduce the gap between the scientific community and society, encouraging the different stakeholders (civil society entities, educational community, the scientific community, policy makers, and the business and industrial sector) to work together throughout the research and innovation process and to interact in the construction of a research agenda. In this sense, studies on gender and altmetrics can inform efforts to promote gender equality in academia and scholarly communication. By identifying areas where gender disparities in altmetric attention are particularly pronounced, researchers and stakeholders can work to develop interventions and strategies to promote greater gender equity in the dissemination and impact of scholarly work.

Altmetrics data (Altmetric, 2020) have the advantage of measuring different types of impact beyond academic citations. They also have the potential to capture earlier evidence of impact. This is useful for self-assessment. However, researchers' social attention should be used with caution, as it may provide a partial and biased view of all types of social impact, in addition to the fact that it does not distinguish between positive and negative impact. Altmetrics have their limitations, particularly in their reliance on social media activity. This can lead to a bias against research that is more likely to be shared on social media, such as studies with catchy titles or controversial findings. In addition, while altmetrics scores provide a snapshot of the online attention a study has received, they lack context (Fang et al., 2022). For example, a study may receive a high number of tweets, but it is unclear whether those tweets express positive or negative sentiment.

Altmetrics can be influenced by many factors, including the visibility and dissemination practices specific to each discipline. The diversity of altmetrics sources and platforms also contributes to differences in social attention across disciplines. It is therefore crucial to interpret altmetrics in the context of each field and to consider them alongside traditional metrics in order to gain a comprehensive understanding of the attention and impact of research across diverse disciplines.

Finally, a positive aspect of the rankings proposed in this paper (i.e., measures of social attention) is that they make it possible to compare completely different fields of research because they standardize the production. Nevertheless, pointing out differences by observing the rankings and testing their differences is only a first approximation to the problem, and more evidence is needed through other statistical techniques. Moreover, for future research, it would be desirable to reproduce the same analysis in relative terms, so that instead of talking about the 250 most influential researchers, it could be analyzed, for example, the 10% most influential researchers, which is not possible with the available dataset. Given that the scientific fields may vary in size, performing the study in relative terms could thus check the robustness of our results and, if necessary, reduce any potential bias.

**Supplementary Material**

*Supplementary Table 1. Spearman's correlation coefficients between the different rankings, by altmetrics*

|  |  | Ranking #1 | | | | Ranking #2 | | | | Ranking #3 | | | |
|---|---|---|---|---|---|---|---|---|---|---|---|---|---|
|  |  | News | Policy | Twitter | Wiki | News | Policy | Twitter | Wiki | News | Policy | Twitter | Wiki |
| Ranking #1 | News | 1 | | | | | | | | | | | |
|  | Policy | 0.01 | 1 | | | | | | | | | | |
|  | Twitter | -0.00 | -0.06**** | 1 | | | | | | | | | |
|  | Wiki | 0.15**** | 0.06**** | -0.07**** | 1 | | | | | | | | |
| Ranking #2 | News | 0.73**** | 0.02 | 0.04 | 0.20**** | 1 | | | | | | | |
|  | Policy | 0.01 | 0.99**** | -0.06*** | 0.07**** | 0.02 | 1 | | | | | | |
|  | Twitter | 0.34**** | 0.14**** | 0.36**** | 0.19**** | 0.36**** | 0.14**** | 1 | | | | | |
|  | Wiki | 0.15**** | 0.07**** | -0.07**** | 0.99**** | 0.20**** | 0.07**** | 0.19**** | 1 | | | | |
| Ranking #3 | News | 0.17**** | -0.00 | 0.11**** | 0.14**** | 0.87**** | -0.00 | 0.23**** | 0.14**** | 1 | | | |
|  | Policy | 0.09 | 0.22**** | 0.12*** | 0.16**** | 0.16**** | 0.58**** | 0.19**** | 0.18**** | 0.14**** | 1 | | |
|  | Twitter | 0.35**** | 0.14**** | 0.31**** | 0.19**** | 0.37**** | 0.14**** | 0.99**** | 0.19**** | 0.23**** | 0.19**** | 1 | |
|  | Wiki | 0.19**** | 0.01 | -0.02 | 0.32**** | 0.21**** | 0.02 | 0.23**** | 0.61**** | 0.14**** | 0.29**** | 0.24**** | 1 |

*Note: Statistically significant at 0.1% (****); 1% (***); or 5% (**). Wiki = Wikipedia*

*Supplementary Figure 1. Distribution of the number of publications with mentions in any of the altmetrics, and the number of publications with some mention in News, Twitter, Wikipedia, and Policy, by gender (Female and Male) and field*

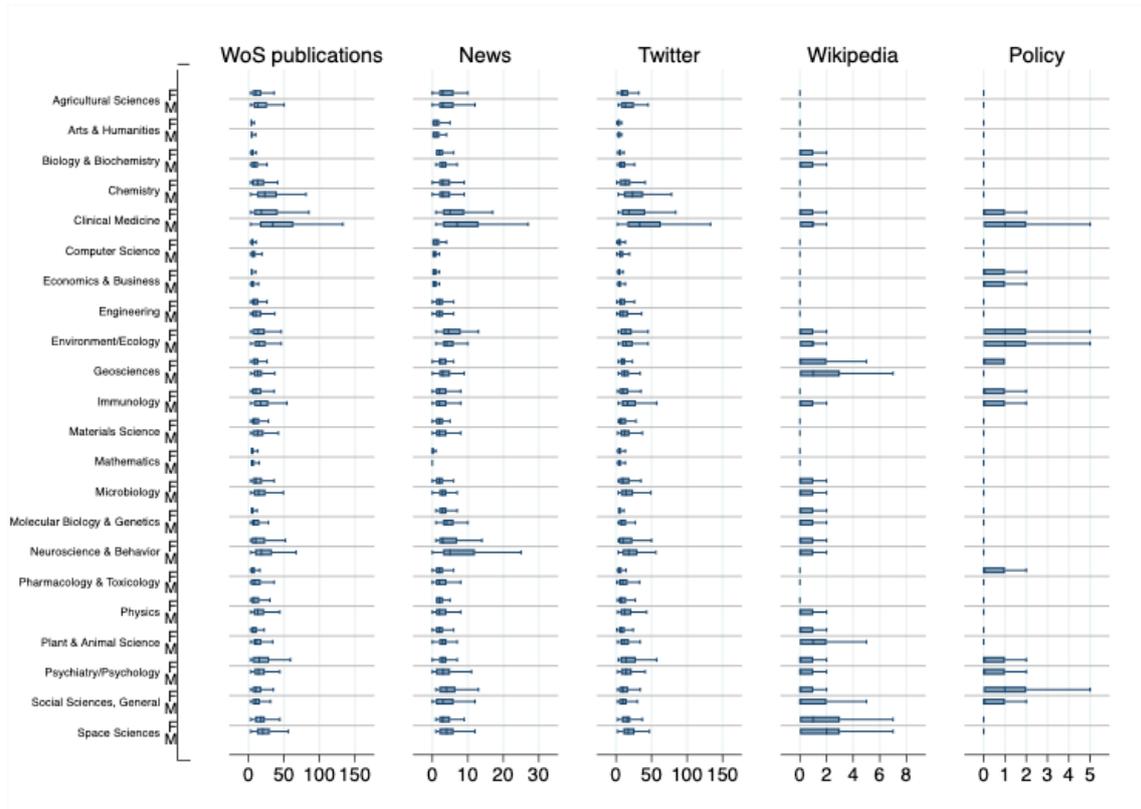

*Supplementary Figure 2. Distribution of the number of publications with mentions in any of the altmetrics, and the number of mentions in News, Twitter, Wikipedia, and Policy, by gender (Female and Male) and field*

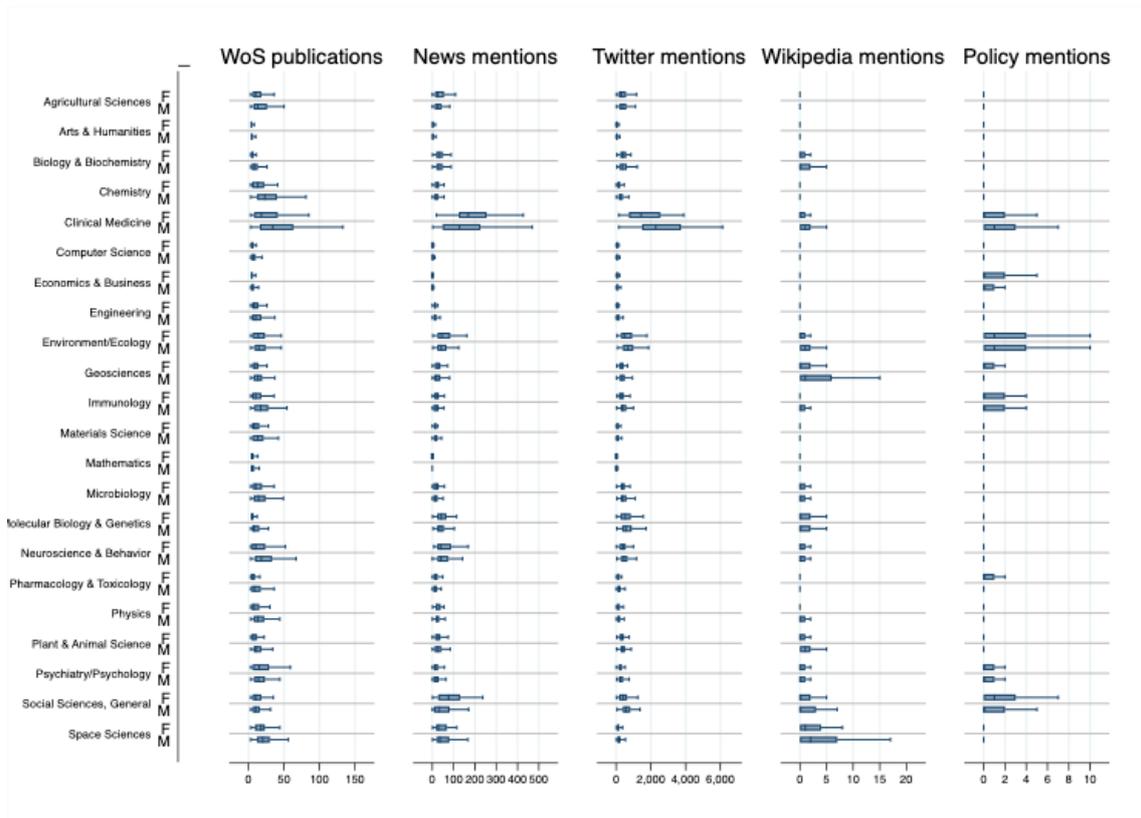